\documentclass[12pt]{iopart}

\usepackage[dvips]{graphicx}

\begin{document}

\title{Gravitational radiation from the $r$-mode instability}

\author{Benjamin J. Owen\dag, and Lee Lindblom\ddag}

\address{\dag Department of Physics, University of
Wisconsin-Milwaukee, P.O. Box 413, Milwaukee, WI 53201 USA}

\address{\ddag Theoretical Astrophysics 130-33, California Institute
of Technology, \hfill \break Pasadena, CA 91125 USA}

\eads{\mailto{owen@gravity.phys.uwm.edu},
\mailto{lindblom@tapir.caltech.edu}}

\begin{abstract}  The instability in the $r$-modes of rotating neutron
stars can (in principle) emit substantial amounts of gravitational
radiation (GR) which might be detectable by LIGO and similar
detectors.  Estimates are given here of the detectability of this GR
based the non-linear simulations of the $r$-mode instability by
Lindblom, Tohline and Vallisneri.  The burst of GR produced by the
instability in the rapidly rotating 1.4$\,M_\odot$ neutron star in
this simulation is fairly monochromatic with frequency near $960\,$Hz
and duration about $100\,$s.  A simple analytical expression is
derived here for the optimal $S/N$ for detecting the GR from this type
of source.  For an object located at a distance of $20\,$Mpc
we estimate the optimal $S/N$ to be in the range $1.2 \sim 12.0$
depending on the LIGO II configuration.

\end{abstract}
\maketitle

\section{Introduction}

This paper contains a portion of the material presented by Lee
Lindblom in the talk ``Relativistic Instabilities in Compact Stars''
at the Fourth Amaldi Conference on Gravitational Waves.  Most of the
material presented in that talk has already been published, and so we
limit our discussion here to that material which has not previously
appeared in print.  We direct those readers who might be interested in
the wider range of subjects covered in the talk to the recent review
Lindblom (2001), and the recent papers Lindblom, Tohline and Vallisneri
(2001b) and Lindblom and Owen (2001).

Gravitational radiation (GR) is a de-stabilizing force on the
$r$-modes of rotating neutron stars (Andersson 1998, Friedman and
Morsink 1998).  The first estimates of the timescales associated with
this instability (Lindblom, Owen and Morsink 1998) showed that the GR
driving force is much stronger than the stabilizing effects of the
simplest forms of viscous dissipation in neutron stars.  As a
consequence there has been a great deal of interest in this
instability as a potential source of observable GR, and as a mechanism
for removing angular momentum from rapidly rotating neutron stars.
During the past several years the effects of a number of additional
dissipation mechanisms on the $r$-mode instability have been studied
in some detail.  A number of these mechanisms are much more effective
in suppressing the $r$-mode instability than the simple viscosity
considered in the initial estimates.  In particular the effects of a
solid crust (Bildsten and Ushomirsky 2000, Lindblom, Owen and
Ushomirsky 2000, Wu, Matzner and Arras 2001), the effects of magnetic
fields (Rezzolla, \etal 2001a, 2001b, Mendell 2001), the non-linear
effects of mode-mode coupling (Schenk, \etal 2001), and the effects of
hyperon bulk viscosity (Jones 2001a, 2001b, Lindblom and Owen 2001)
make it appear less likely that the GR instability in the $r$-modes
will play an interesting role in astrophysics.  However, at the
present time none of these mechanisms is understood well enough for us
to conclude absolutely that the $r$-mode instability will never play a
role in any neutron stars.  Thus for the purposes of the present
paper, we assume that the instability will occur in some rapidly
rotating neutron stars.  Our aim here is to present the best estimate
of the GR that might be emitted during such an instability.  To do
this we analyze the GR emitted by the best currently available
numerical simulation of the non-linear evolution of the $r$-mode
instability.  This paper is an update of the initial estimates of the
GR emitted by the $r$-mode instability given by Owen, \etal (1998).

\section{{\it r}-Mode Evolution Model}

\begin{figure}
\begin{center}
\vskip 0.6cm
\includegraphics[height=2.0in]{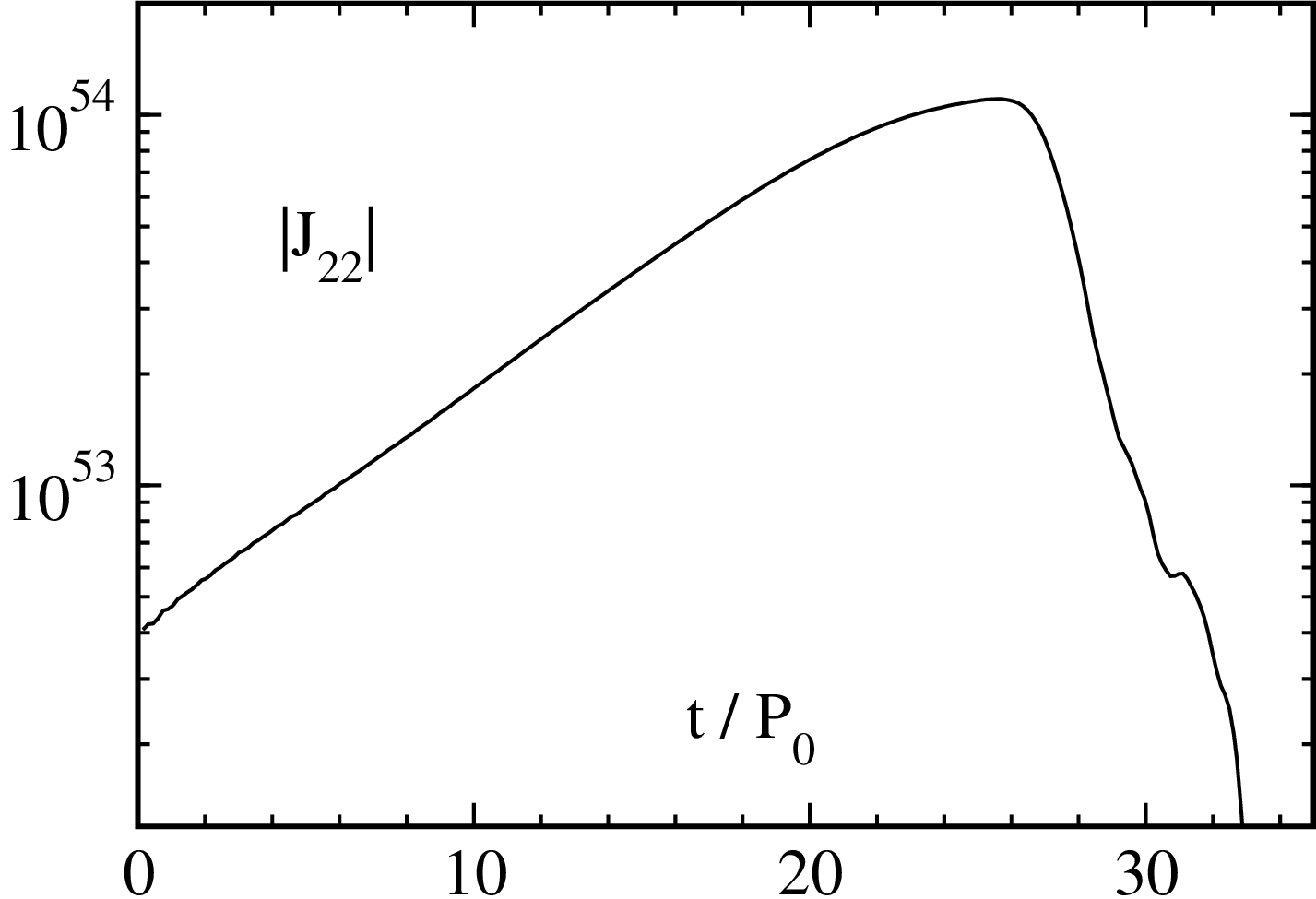}
\caption{\label{fig1}Evolution of the current quadrupole moment
$J_{22}$ (in cgs units) from the numerical simulation of the GR driven
growth of the $m=2$ $r$-mode.}
\end{center}
\end{figure}

We base our estimates of the GR produced by the non-linear evolution
of an unstable $r$-mode, on the numerical simulations by Lindblom,
Tohline and Vallisneri (2001a, 2001b).  The evolution of a small
amplitude $m=2$ $r$-mode subject to the current-quadrupole GR reaction
force was studied numerically in a $1.4M_\odot$ polytropic neutron star
rotating at about 95\% of its breakup angular velocity.  Figure~\ref{fig1}
illustrates the evolution of the current quadrupole moment $|J_{22}|$
(in cgs units) of this model as the simulation evolves.  The current
quadrupole moment is defined as
\begin{equation}
J_{22}=\int \rho r^2 \vec v \cdot \vec Y^{B*}_{22} d^{\,3}x,
\end{equation}
where $\rho$ and $\vec v$ are the density and velocity of the fluid in
the star, and $\vec Y^{B}_{22}$ is the magnetic type vector spherical
harmonic (Thorne 1980). The time displayed along the horizontal axis
of Fig.~\ref{fig1} is given in units of the initial rotation period of
the star: $P_0=1.18\,$ms.  During the first part of the evolution GR
drives the growth of the $r$-mode, leading to the exponential growth in
$J_{22}$ illustrated here.  Once the amplitude of the mode becomes
sufficiently large, however, non-linear hydrodynamic processes also
become important.  These lead to large surface waves which break and
shock (see Lindblom, Tohline and Vallisneri 2001a, 2001b).  The
dissipation in these shocks damps the $r$-mode and leads to the rapid
decrease in $J_{22}$ following its peak.

The timescale of the GR instability in the $r$-modes is much much
longer than the hydrodynamic timescale.  To perform the numerical
simulation illustrated in Fig.~\ref{fig1} it was necessary to increase
artificially the strength of the GR driving force so that the
instability could proceed more rapidly.  Fortunately tests have shown
(see Lindblom, Tohline and Vallisneri 2001b) that the maximum
amplitude which the $r$-mode achieves (and hence the maximum value of
$J_{22}$) is relatively insensitive to the strength of the GR driving
force.  The amplitude grows until the velocities in the mode reach
some critical value before the waves break and shock.  The value of
this critical velocity appears to be relatively insensitive to how
rapidly the fluid velocity is increased to this critical level.

To determine the GR signal that would be emitted by the growth of an
unstable $r$-mode, it is necessary to re-scale the time in the
simulation.  The early part of the simulation has been designed to
proceed more rapidly than the physical case by the factor
$\tau_{GR}({\rm physical})/\tau_{GR}({\rm simulation})$, where
$\tau_{GR}$ represents the GR growth timescale of the instability.
This ratio has the value 4488 in the simulation used here.  Thus the
early exponential growth phase of a physical $r$-mode evolution will
last longer than this phase of the simulation by this factor.  The
late stages of the simulation are dominated by hydrodynamic forces
which damp the $r$-mode within a few rotation periods.  This final
stage of the evolution therefore proceeds at the same rate in both the
physical and the simulation cases.  We (somewhat artificially) choose
the dividing time between the early and late stages of the evolution
to be the time at which the value of $|J_{22}|$ is maximum.

\begin{figure}
\begin{center}
\vskip 0.6cm
\includegraphics[height=2.0in]{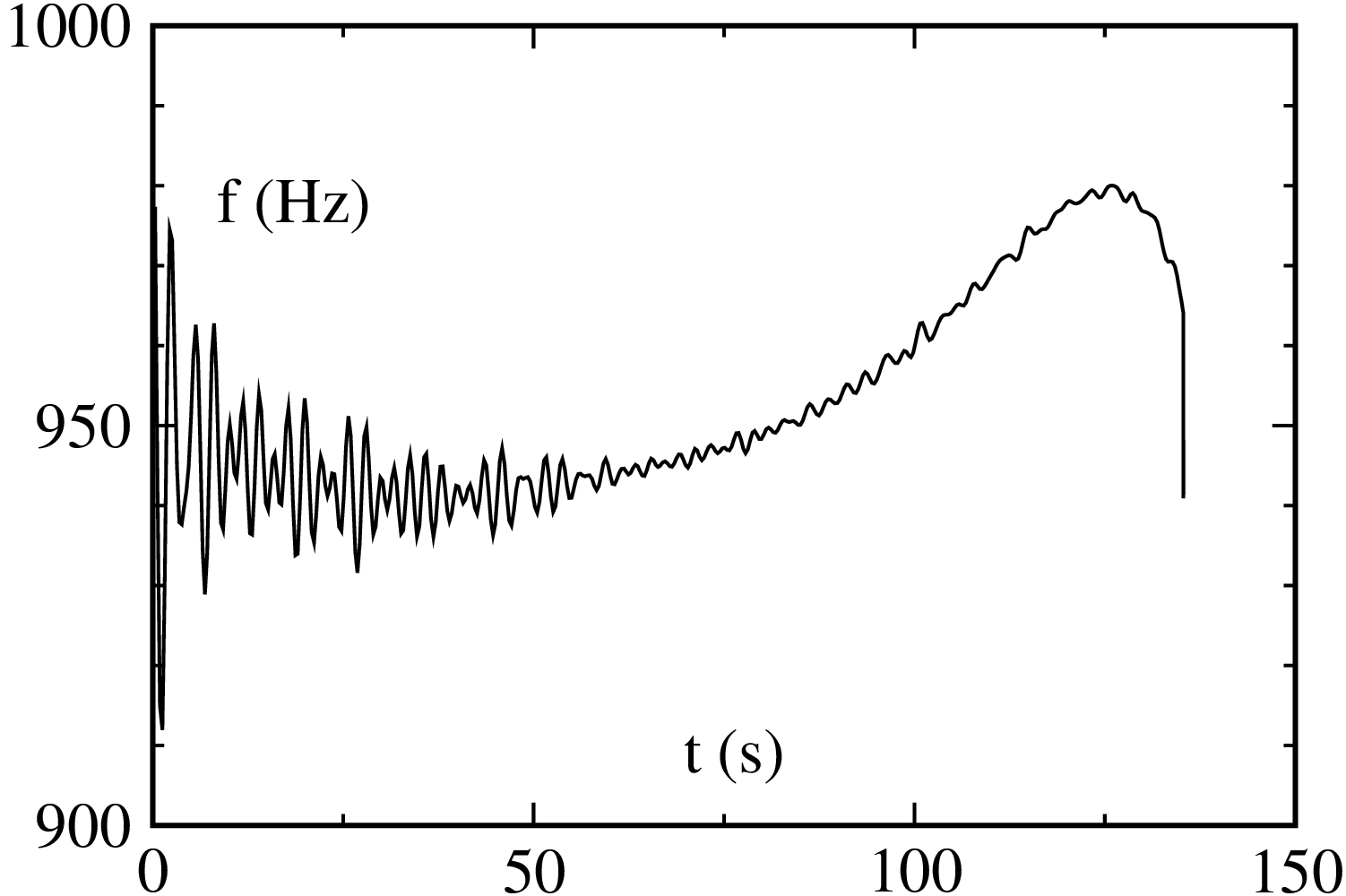}
\caption{\label{fig2}Evolution of the ``frequency'' of the $r$-mode.
The time parameter used here (in seconds) is scaled from the simulation to
reflect the physical case until the formation of shocks. Noise early in the
simulation is due to contamination of the current quadrupole $J_{22}$ by
other modes.}
\end{center}
\end{figure}

The time dependence of $J_{22}$ is found in the simulation to be
nearly sinusoidal with a time dependent amplitude,
$J_{22}=|J_{22}|e^{i\psi(t)}$, (with $d\psi/dt=\omega\approx$ constant)
because $J_{22}$ is dominated by the unstable $r$-mode in this case.
The time dependence of $|J_{22}|$ is illustrated in Fig.~\ref{fig1}.
The frequency of the sinusoidal time dependence is conveniently
determined numerically using the formula
\begin{equation}
\omega = - {1\over |J_{22}|}\left|{d J_{22}\over dt}\right|.
\end{equation}
This approximation introduces errors of order
$(\omega\tau_{GR})^{-1}\approx 2\%$ in our simulation.  Figure~\ref{fig2}
illustrates the evolution of the frequency, $f=-2\pi\omega$,
determined numerically in this way.  Here we plot the frequency as
a function of the re-scaled physical time (in seconds).  We note that
the frequency changes by only a few percent during the course of the
evolution in which about 40\% of the angular momentum of the star
is radiated away as GR.

\begin{figure}
\begin{center}
\vskip 0.6cm
\includegraphics[height=2.0in]{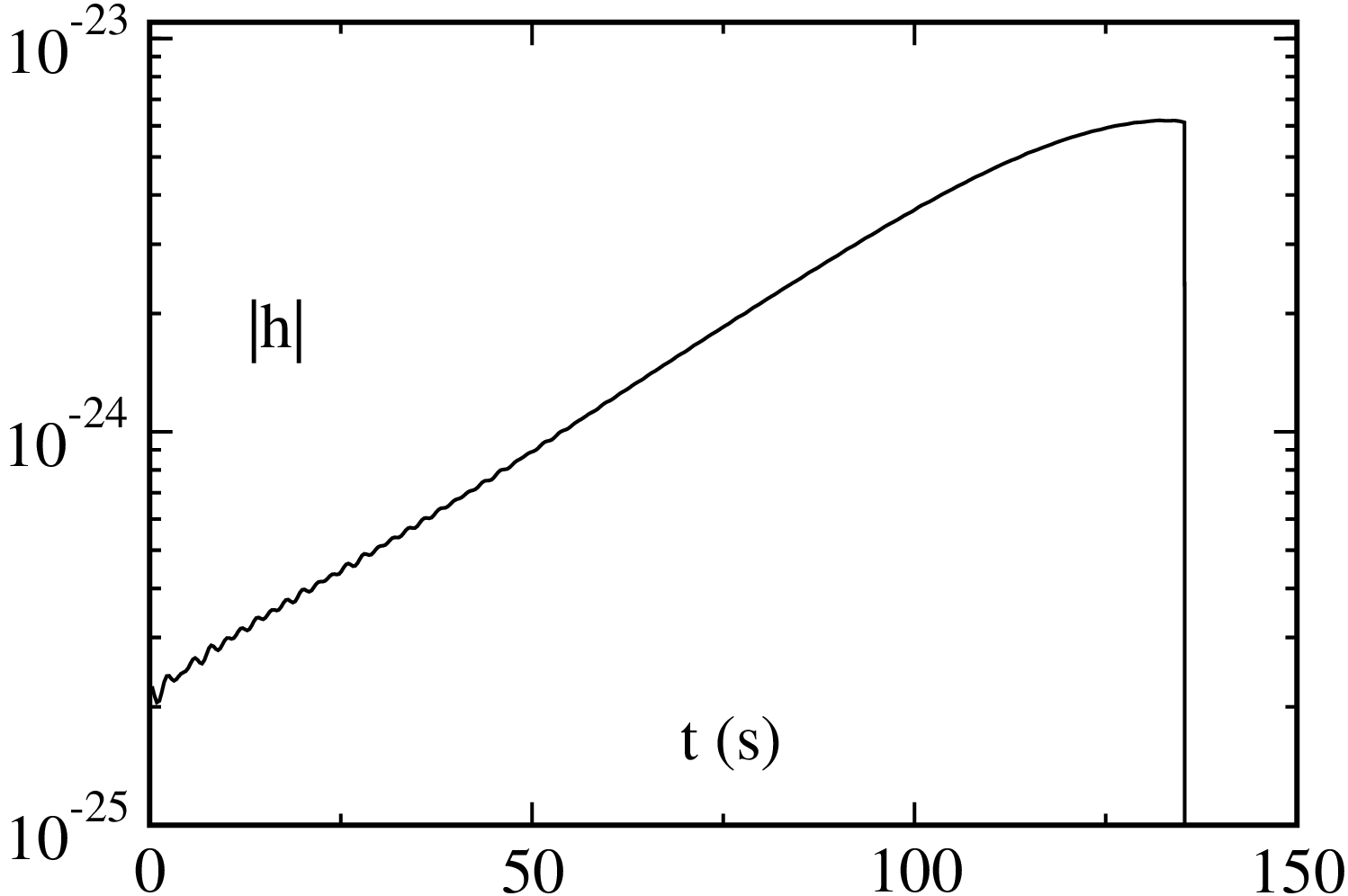}
\caption{\label{fig3}Evolution of the angle averaged dimensionless
gravitational wave amplitude $|h(t)|$ for a source located at a
distance $D=20\,$Mpc.  The time parameter used here (in seconds) is scaled
from the simulation to reflect the physical case until the formation of
shocks.}
\end{center}
\end{figure}

In general relativity theory an isolated object with time dependent
current quadrupole moment radiates GR.  The expression for the
dimensionless GR amplitude $h$ (averaged over possible source and detector
orientations) is given by
\begin{equation}
h(t) = \frac{16}{15} \sqrt{\frac{2\pi}{5}}\frac{\omega^2 G
|J_{22}|}{c^5 D}e^{i\psi(t)},\label{e:ht}
\end{equation}
where $G$ and $c$ are Newton's constant and the speed of light, and
$D$ is the distance to the source.  Figure~\ref{fig3} illustrates the
time dependence of this dimensionless GR amplitude as a function of
time as determined by the simulation.  Since the frequency of these
$r$-mode oscillations is essentially constant, the GR amplitude just
grows in proportion to $J_{22}$.  Once the amplitude peaks, non-linear
hydrodynamic forces (i.e. shocks) quickly damp the mode and this is
reflected in the sharp drop in $h$ (as a function of physical time)
illustrated in Fig.~\ref{fig3}.  We have chosen $D=20\,$Mpc, the
distance to the Virgo Cluster of galaxies, for the purposes of the
illustration.  Optimistic estimates of the event rate for these
objects suggests that the nearest events expected during the
time frame of the LIGO observations will be at this distance.

\section{Optimal {\it S/N}}

\begin{figure}
\begin{center}
\vskip 0.6cm
\includegraphics[height=2.0in]{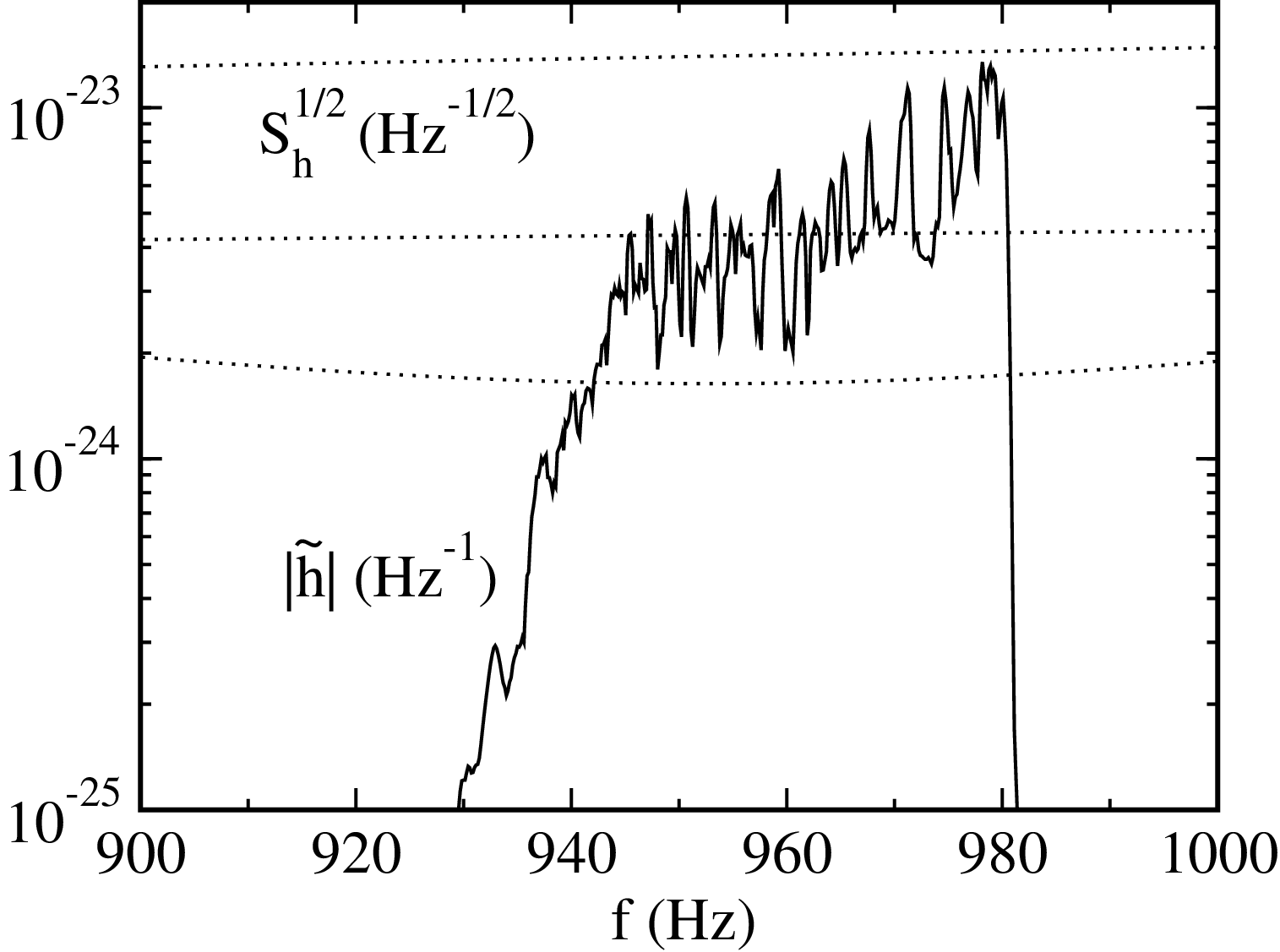}
\caption{\label{fig4}Fourier transform of the gravitational wave
amplitude illustrated in Fig.~\ref{fig3}, $|\tilde h(f)|$.  Also shown
are the noise power spectral density $S^{1/2}_h$ for the three
potential LIGO II configurations used in our estimates of the optimal
$S/N$.}
\end{center}
\end{figure}

The optimal signal to noise ratio ($S/N$) for detecting a GR signal
can be achieved with the use of an optimal filter, consisting of a
template that matches the waveform of the signal.  Let $\tilde h$
denote the Fourier transform of the GR signal:
\begin{equation}
\tilde h(f)=\int_{-\infty}^{\infty} h(t) e^{-2\pi i f t} dt.
\end{equation}
The discrete Fourier transform of the gravitational wave signal from
our numerical simulation is shown in Fig.~\ref{fig4}.  Here we have
smoothed the resulting $|\tilde h(f)|^2$ with a windowing function of
width $0.5\,$Hz.  We see that the spectrum of the gravitational
radiation is confined essentially to the band $940\leq f\leq 980\,$Hz,
which corresponds well with the evolution of the frequency of the
$r$-mode as shown in Fig.~\ref{fig2}.  For a complex signal $h(t)$ the
optimal value of $S/N$ is given by
\begin{equation}
\left(\frac{S}{N}\right)^2=2\int_0^\infty\frac{|\tilde h(f)|^2 df}{S_h(f)},
\label{e:sn}
\end{equation}
where $S_h(f)$ is the one-sided power spectral density of detector
noise.  (For a real signal the 2 becomes a 4.)  We approximate
$S_h(f)$ for frequencies near $960\,$Hz by the
Taylor expansion
\begin{equation}
S_h(f)\approx S+S'\Delta f + 0.5S''(\Delta f)^2,\label{e:sh}
\end{equation}
where $\Delta f = f-960\,$Hz, and the three constants $S$, $S'$ and
$S''$ are listed in Table~\ref{table1} for three plausible LIGO II
design options (see Buonanno and Chen 2001).  Using these values for
$S_h$ we find $S/N\approx 1.2$, 4.0 and 10.4 by performing the
integral in Eq.~(\ref{e:sn}) numerically for these three possible LIGO
II configurations.
\begin{table}
\caption{\label{table1}Coefficients for the Taylor expansion of the
one-sided power spectral density $S_h$ at $f=960\,$Hz for
three plausible LIGO II configurations.}
\begin{indented}
\item[]\begin{tabular}{@{}lr@{}lr@{}lr@{}l}
\br
Configuration&&$\quad S$&&$\quad S'$&&$\quad S''$\\
\mr
NS--NS Optimized&$2.$&$0\times 10^{-46}\,$Hz${}^{-1}$ &$5.$&$0\times 10^{-49}$
Hz${}^{-2}$&$5.$&$6\times 10^{-52}\,$Hz${}^{-3}$\\
Broadband&$1.$&$9\times 10^{-47}\,$Hz${}^{-1}$ &$2.$&$2\times 10^{-50}$
Hz${}^{-2}$&$2.$&$6\times 10^{-53}\,$Hz${}^{-3}$\\
Narrowband&$2.$&$7\times 10^{-48}\,$Hz${}^{-1}$ &$6.$&$0\times 10^{-51}$
Hz${}^{-2}$&$8.$&$1\times 10^{-52}\,$Hz${}^{-3}$\\
\br
\end{tabular}
\end{indented}
\end{table}

We can also make an analytical estimate of the optimal $S/N$ using an
extension of a very general argument given originally by Blandford
(1984, unpublished).  For GR emitted by a multipole with azimuthal
quantum number $m$ the angular momentum loss rate is
\begin{equation}
\frac{dJ}{dt}=-\frac{5m\pi c^3}{2G}f D^2 |h(t)|^2,\label{e:djdt}
\end{equation}
where $|h(t)|^2$ has been averaged over all possible orientations of source
and detector (and over several wave periods for the case of a real signal).
When a function such as $h(t)$ involves a rapid oscillation together with a
much slower evolution of its amplitude and frequency, the Fourier transform
$\tilde h(f)$ is well approximated by the stationary phase approximation,
\begin{equation}
|h(t)|^2 =|\tilde h(f)|^2\left|\frac{df}{dt}\right|\label{e:sphase}
\end{equation}
for a complex signal $h(t)$. (For a real signal, the left-hand side is
averaged over several periods and the right-hand side is multiplied by 2.)
Since the angle-averaging affects both sides of Eq.~(\ref{e:sphase})
equally, we can use Eqs.~(\ref{e:djdt}) and (\ref{e:sphase}) to re-express
$S/N$ for a source at an ``average'' orientation from Eq.~(\ref{e:sn}) as
\begin{equation}
\left(\frac{S}{N}\right)^2=-\frac{4G}{5m\pi c^3 D^2}\int
\frac{dJ}{fS_h(f)}.\label{e:sn2}
\end{equation}
For GR sources such as the $r$-mode evolution the frequency of the
radiation emitted is nearly constant, so $fS_h(f)$ can be treated as
being essentially constant in Eq.~(\ref{e:sn2}).  Thus, this integral
becomes just the total amount of angular momentum $\left|\Delta
J\right|$ radiated away as GR:
\begin{equation}
\left(\frac{S}{N}\right)^2=\frac{4G}{5m\pi c^3 D^2}
\frac{\left|\Delta J\right|}{fS_h(f)}.\label{e:sn3}
\end{equation}
The total amount of angular momentum radiated away as GR in the
simulation was $|\Delta J|\approx 4.5\times 10^{48}$ in cgs units.
Using this value in Eq.~(\ref{e:sn3}) together with the values of
$S_h$ from Eq.~(\ref{e:sh}) and Table~\ref{table1}, we find
$S/N\approx 1.4$, 4.5, and 12.0 in good agreement with our previous
estimate based on the discrete Fourier transform and direct
integration of the simulation waveform.  We note that this argument
can easily be extended to signals composed of multiple harmonics and
multipoles so
long as their frequencies are well-defined: Time-averaging of
Eq.~(\ref{e:sphase}) eliminates cross-terms, so Eq.~(\ref{e:sn3}) is simply
summed over each harmonic and $m$. We can also extend the argument to
non-monotonically evolving frequencies by using Eq.~(\ref{e:sphase})
to express Eq.~(\ref{e:sn}) as an integral over time and dividing
$f(t)$ into piecewise monotonic parts.

It is not unreasonable to think that a realistic data analysis strategy
could come within a factor of 2 of the optimal $S/N$. Due to the complexity
of the physics involved it seems unlikely that matched filtering will ever
be a viable option. However, cross-correlation of the output of two aligned
interferometers (LIGO Hanford and LIGO Livingston) can in principle achieve
$1/\sqrt{2}$ of optimal $S/N$ if the detectors' noise is of comparable
strength and uncorrelated---including non-Gaussian noise bursts (Anderson
\etal 2001). This strategy relies on the supernova associated with the
$r$-mode having been observed optically, allowing the appropriate time delay
to be inserted between the interferometer data streams. Narrowing the search
to a few minutes after the supernova (instead of continuous operation) also
has the effect of greatly reducing the $S/N$ threshold for detection with
reasonable false alarm statistics. Thus a cross-correlation with
$S/N\approx4$ might be considered enough for detection, implying a
realistically detectable distance for $r$-modes of 5~Mpc for even the least
optimal (for this type of source) LIGO II configuration---and up to 50~Mpc
for the most optimal.

\ack We thank Joel Tohline and Michele Vallisneri for allowing us to use the
results of the non-linear $r$-mode simulations, Alessandra Buonanno for
providing us with the latest LIGO II noise curves, and Jolien Creighton for
conversations on data analysis.  We also thank Kip Thorne for helpful
comments and suggestions.  This research was supported by NSF grants
PHY-9796079, PHY-0071028, PHY-0079683, and PHY-0099568, and NASA grants
NAG5-4093 and NAG5-10707.

\References

\item[] Anderson W G, Brady P R, Creighton J D E and Flanagan \'E \'E 2001
{\it Phys. Rev. D} {\bf 63} 042003

\item[] Andersson N 1998 {\it Astroph. J.} {\bf 502} 708

\item[] Bildsten L and Ushomirsky G 2000 {\it Astroph. J.} {\bf 529} L33

\item[] Buonanno A and Chen Y 2001 {\it Phys. Rev. D} {\bf 64} 042006

\item[] Friedman J L and Morsink S M 1998 {\it Astroph. J.} {\bf 502} 714

\item[] Jones P B 2001a \PRL {\bf 86} 1384

\item[] Jones P B 2001b {\it Phys. Rev. D} {\bf 64} 084003

\item[] Lindblom L 2001 in {\it Gravitational Waves: A Challenge to
Theoretical Astrophysics}, ed V Ferrari \etal
(Trieste: ICTP Lecture Notes, Vol 3) pp~257--275;
http://www.ictp.trieste.it

\item[] Lindblom L and Owen B J 2001 {\it Phys. Rev. D} (submitted);
astro-ph/0110558

\item[] Lindblom L, Owen B J and Morsink S M 1998 \PRL {\bf 80} 4843

\item[] Lindblom L, Owen B J and Ushomirsky G 2000 {\it Phys. Rev. D},
{\bf 62}, 084030

\item[] Lindblom L, Tohline J E and Vallisneri M 2001a \PRL
{\bf 86} 1152

\item[] Lindblom L, Tohline J E and Vallisneri M 2001b {\it Phys. Rev. D}
(submitted); astro-ph/0109352

\item[] Mendell G 2001 {\it Phys. Rev. D} {\bf 64} 044009

\item[] Owen B J, Lindblom L, Cutler C, Schutz B F, Vecchio A and Andersson N
1998 {\it Phys. Rev. D} {\bf 58} 084020

\item[] Rezzolla L, Lamb F L, Markovic D and Shapiro S 2001a {\it
Phys. Rev. D} {\bf 64} 104013

\item[] Rezzolla L, Lamb F L, Markovic D and Shapiro S 2001b {\it
Phys. Rev. D} {\bf 64} 104014

\item[] Schenk A K, Arras P, Flanagan E E, Teukolsky S A, Wasserman I
gr-qc/ 0101092

\item[] Thorne K 1980 \RMP {\bf 52} 299

\item[] Wu Y, Matzner C D and Arras P 2001 {\it Astrophys. J} {\bf 549} 1011

\endrefs

\end{document}